# *Suspended Monolayer Graphene under True Uniaxial Deformation*

**IOANNIS POLYZOS[1*], MASSIMILIANO BIANCHI[2], LAURA RIZZI[2], EMMANUEL KOUKARAS[1], JOHN PARTHENIOS[1], KONSTANTINOS PAPAGELIS[1,3], ROMAN SORDAN[2] AND COSTAS GALIOTIS[1,4*]**

[1]Institute of Chemical Engineering Sciences, Foundation of Research and Technology-Hellas (FORTH/ICE-HT), Patras, Greece

[2]L-NESS, Department of Physics, Politecnico di Milano, Polo di Como, Via Anzani 42, 22100 Como, Italy

[3]Department of Materials Science, University of Patras, Greece

[4]Department of Chemical Engineering, University of Patras, Greece

*Corresponding authors*: c.galiotis@iceht.forth.gr, ipolyzos@iceht.forth.gr

# *Abstract*

**2D crystals, such as graphene, exhibit the highest strength and stiffness than any other known man-made or natural material. So far, this assertion is primarily based on modelling predictions and on bending experiments in combination with pertinent modelling. True uniaxial loading of suspended graphene is not easy to accomplish; however such an experiment is of paramount importance in order to assess the intrinsic properties of graphene without the influence of an underlying substrate. In this work we report on uniaxial tension of graphene up to moderate strains of ~0.8%. This has been made possible by sandwiching the graphene flake between two polymethylmethacrylate (PMMA) layers and by suspending its central part by the removal of a section of PMMA with e-beam lithography. True uniaxial deformation is confirmed by the measured large phonon shifts with strain with Raman spectroscopy and the indication of lateral buckling (similar to what is observed for thin macroscopic membranes under tension). Finally, we also report on how the stress is transferred to the suspended specimen through the adhesive grips and determine the value of interfacial shear stress that is required for efficient axial loading in such a system.**

# *Introduction*

Graphene consists of a two-dimensional (2D) sheet of covalently bonded carbon atoms and forms the basis of both one-dimensional carbon nanotubes, three-dimensional graphite but also important commercial products, e.g., polycrystalline carbon (graphite) fibres. [1-3] Due to its extraordinary properties, a continuously increasing amount of applications is emerging in fields like nanoelectronics[4,5], micro- and nanomechanical systems[6,7], sensors[8], optoelectronics, photonics[9], composite materials[10] etc. As a single defect-free molecule, graphene is predicted to have an intrinsic tensile strength (130 GPa) higher than any other known material[11] and tensile stiffness (1TPa) similar to values measured for graphite.

The mechanical stretching of freely-suspended graphene is of paramount importance for understanding the mechanical behaviour of the material free of doping or other unwanted influences by the underlying substrate (e.g. roughness, impurities etc). Zabel *et al*[12] used graphene bubbles to study graphene under biaxial (e.g., isotropic) strain and derived the Grüneisen parameters through Raman spectroscopy (RS). Lee *et al* [13] loaded graphene in biaxial tension by the simple bending of a tiny flake by an indenter on an AFM set-up. By considering graphene as a clamped circular membrane made by an isotropic material of zero bending stiffness, they converted the bending force vs deflection curve to an "axial" stress-strain curve. This way they managed to confirm the extreme stiffness of graphene of 1 TPa and provided an indication of the breaking strength of graphene of 42 N m$^{-1}$ (or 130 GPa considering graphene thickness as 0.335 nm).

A great deal of work has already been conducted on supported single graphene flakes, which have been subjected to axial tension and compression using beam-type loading systems developed in the early nineties.[14] These experiments[8,15-19] confirmed the extreme stiffness of graphene of about 1 TPa and have provided an estimate of the compression strain to failure of embedded single flakes of approximately -0.6%, regardless of flake geometry. The main technique to monitor the mechanics at the molecular scale is that of laser RS. The Raman spectrum of graphene has only one peak (termed G) corresponding to an $E_{2g}$ phonon in the first order region around 1580 cm$^{-1}$. Defective sp$^2$ bonded carbons also exhibit another peak at 1350 cm$^{-1}$, which is due to the breathing modes of six-atom rings that is activated by an intravalley scattering process and requires a defect for its activation. The 2D peak is the second order of the D peak and is a single peak in monolayer graphene (1LG), whereas it splits into multiple bands in few layer graphenes, reflecting the evolution of the electronic band structure. [20-22] The 2D peak does not require the presence of defects since momentum conservation is obtained by the participation of two phonons with opposite wave vectors ($q$ and $-q$). In all cases due to the anharmonicity of the sp$^2$ bonds, clear shifts to lower (in tension) or higher (in compression) wavenumbers have been observed. For graphene, the actual shifts per strain are quite large and the highest ever observed for any known crystalline material. This high sensitivity makes the Raman technique an ideal monitoring sensor for any strain (or stress) experienced by the investigated material (in our case graphene).[8] The absence of any wavenumber shift indicates unequivocally that the material is not stressed. Particularly, for perfect crystals, such as the 2-dimensional graphene, there is no plastic deformation upon tensile loading that could account for the absence of phonon shifts. In conclusion, for flexed-beam configurations, a linear relationship between Raman frequency and strain has been obtained in tension up to maximum strains of the order of 2%. However, due to restrictions of the flexed-beam configuration, the supported graphene monolayers cannot be strained to higher than 2% strain.

The reported experiments above, correspond to either biaxial loading or uniaxial measurements on supported specimens. Attempts to load uniaxially suspended graphene monolayers to fracture are scarce. The reason for this is the difficulties associated with the suspension of graphene flakes of micrometre dimensions over a trench and more importantly the efficient clamping of the specimens. Recently Pérez-Garza et al [23,24] reported tensile experiments on suspended graphene using MEMS as a loading apparatus operating at high temperatures. These authors have attempted to load 3 and 4-layer graphenes, as well as, a monolayer. For the 3LG they reported [3] maximum strains of 12.5% at an applied force of 1.75 mN and more recently they reported maximum strains of 14% for a monolayer at 330 μN. As argued therein such force would only induce a strain of 7.5% on multilayer graphenes in spite of the fact that no major moduli changes in the axial direction are expected for multi-layer graphenes or even graphite. Most importantly the authors did not observe any measurable peak-shifts of Raman spectra for the few layer graphene samples in spite of the fact that, as mentioned also above, for all graphitic materials (graphene, graphite and even carbon fibres) any stress imposition is associated with detectable Raman phonon shifts [8,16,18,19,25]. Even in the case of 1LG the reported shift of the 2D peak [23,24] is about 3 times lower than the expected 2D peak shift in air (~80 cm$^{-1}$/%). Moreover, the reported shift did not seem to emanate from the main 2D feature and therefore its origin is questionable. The low (1LG) or even zero 2D phonon shift with strain is indicative of the fact that the applied strain is either only partially transferred to graphene (1LG) or it is not transferred at all (3LG and 4LG). Another recent experiment reported by Zhang et al[26] on the fracture toughness of CVD bilayer graphene involved the integration of a micromechanical device and a nanoindenter. By moving the nanoindenter tip the authors claim that the specimen was subjected to pure tension by the inclined beams of their device. Axial stress-strain curves of specimens containing cracks are indeed shown for maximum strains as low as 0.3%. The technique certainly represents a solid progress beyond current attempts as it operates at room temperature but although Raman spectra are taken, there is no indication that indeed phonon shifts do occur. As argued above and bearing in mind the complexity of the experiment, such verification is required to ascertain proper axial loading of the specimen.

In the work reported here, a fabrication method is described for the applications of uniaxial strain to suspended graphene. Mechanically exfoliated monolayer graphene sheets were sandwiched between two layers of polymethylmethacrylate (PMMA) and suspended areas were induced by removal of a section of the polymer by e-beam exposure allowing precise control over shape and position (figure 1, see also Methods). As elaborated below, it has been found that the suspended flakes are subjected to a well-defined gradient of strain as a result of the fabrication procedure. By manipulating a Raman microprobe with a nanomover, we have managed to produce accurate strain maps of the suspended flake along and across the strain axis at 100 nm translation steps. Large shifts and splitting of the main peaks of graphene spectrum under a uniaxial strain were recorded in accordance with theoretical predictions. Furthermore, our approach allowed us also to monitor the strains within the portion of graphene embedded ("gripped") into the polymer and to assess, from the obtained strain profiles, the strength of the bonding between graphene and polymer[27]. Finally, the obtained Raman intensity and frequency maps in the lateral direction pointed to the presence of orthogonal wrinkles that can be attributed to the Poisson's contraction in the lateral direction. Based on this assumption the variation of strain within each wrinkle has been identified for the first time and its significance is assessed.

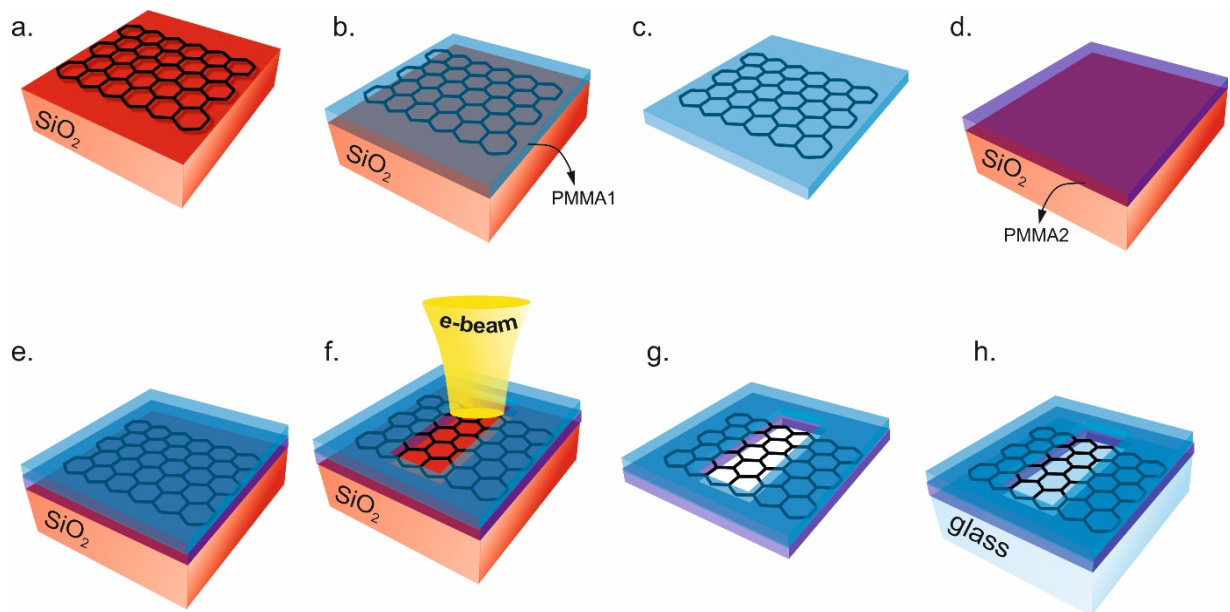

**Figure 1** Schematic of sample preparation: (a)Exfoliation and removal of unwanted flakes by RIE (b)Spin coating of the top PMMA layer (c)Detachment of the top PMMA layer with graphene attached to PMMA (d)Spin coating of the bottom PMMA layer on another chip (e)Detached top PMMA layer with graphene transferred on top of the bottom PMMA layer (f)Windows opened across the graphene flake by e-beam exposure (g)Detachment of the entire structure, h)Transfer of whole assembly to a glass support for ease of handling.

## *Results and Discussion*

In the case examined here, a gradient of axial strain is developed in the suspended graphene as a result of the fabrication method described above, which involved the patterning of windows of micrometer scale in a double-layered PMMA stack. The graphene specimen was sandwiched between the layers and by means of e-beam lithography a certain section of the embedded graphene was fully exposed as shown in Figures 1,2 (see also Methods and Ref [28]). The window area was carefully selected in order to have both suspended graphene and graphene clamped on both sides within the PMMA layers. These layers were prepared by spin coating PMMA resist onto a Si substrate. In this manner, stress is accumulated on the sample, which is normally partially relieved after development. However, due to the large PMMA thickness (~2μm), a residual stress is present upon development. This built-in stress results in the formation of a notable arch upon the two faces of the gap that engulfs graphene. Hence, a gradient of intrinsic uniaxial tensile stresses/ strain is developed across the graphene, which is along the Y-axis of Figure 2.

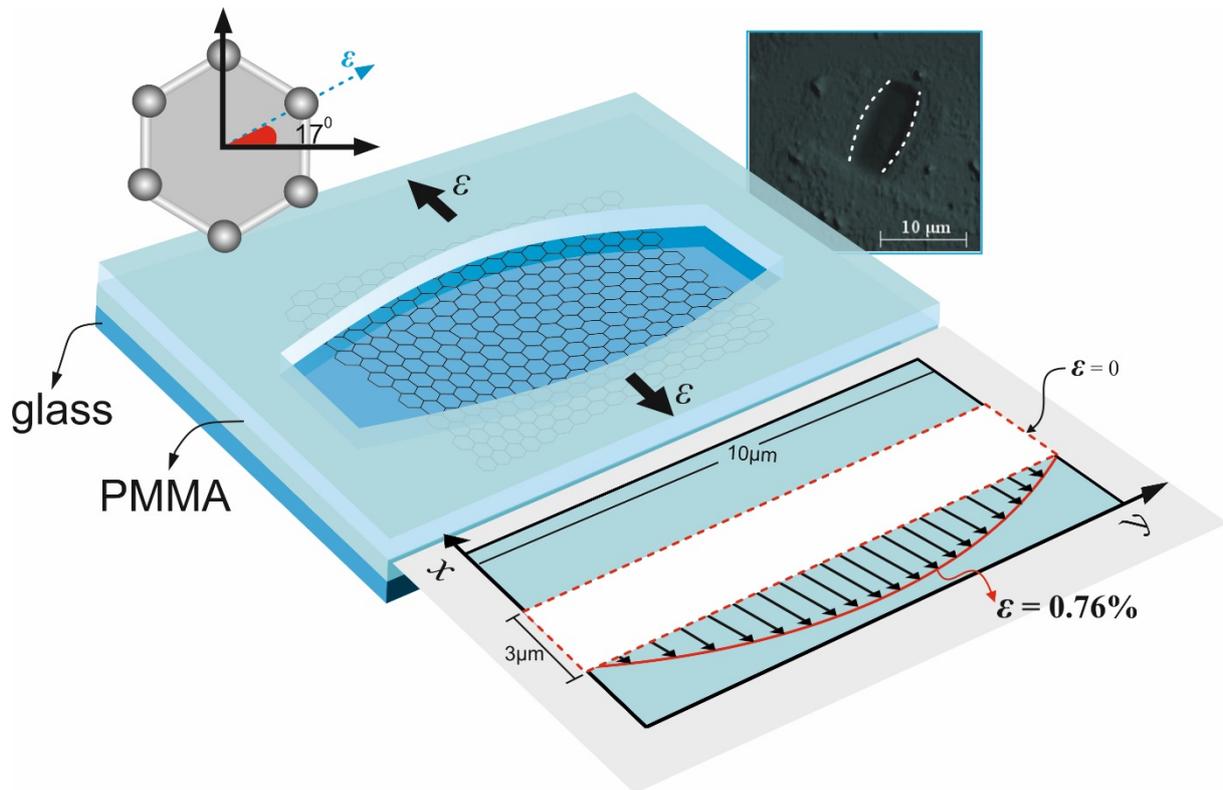

**Figure 2 Schematic of the suspended graphene strain device. Graphene was sandwiched between two PMMA layers and a portion of it was suspended in air by removing a PMMA section using e-beam lithography. The induced intrinsic strain gradient across the flake (y-axis) and the actual strain as derived from the shift of the G peak are shown. The white rectangle inside the x-y coordinate system is the initial window shape prior to the imposition of intrinsic strain. The microphotograph of the sample investigated is shown in the upright corner of the figure. The calculated orientation of graphene with respect to the strain axis is depicted on the upper left corner of the sketch.**

The applied strain values across the specimen can be calculated from microscopic observations of the arch shape and dimensions; however, more accurate values of strain can be obtained by translating the laser beam across the arch and by measuring the phonon shift at each position. In particular, the Raman G-peak shifts according to the following well-established [18] secular equation:

$$\Delta\omega_G^\pm = \Delta\omega_G^h \pm \frac{1}{2}\Delta\omega_G^s = -\omega_G^0 \gamma_G \left(\varepsilon_{ll} + \varepsilon_{tt}\right) \pm \frac{1}{2}\omega_G^0 \beta_G \left(\varepsilon_{ll} - \varepsilon_{tt}\right) \quad (1)$$

where $\Delta\omega_G^h$ and $\Delta\omega_G^s$ are the shifts resulting from the hydrostatic and the shear components of the strain, respectively, $\Delta\omega_G^+$ and $\Delta\omega_G^-$ are the shifts of the $G^+$ and $G^-$ sub-peaks relative to zero strain, $\varepsilon_{ll}$ and $\varepsilon_{tt}$ are the parallel and perpendicular strains, $\gamma_G$=1.99 is the Grüneisen parameter and $\beta_G$=0.99 is the shear deformation potential[18]. The shear strain component is responsible for the splitting of the G peak. The phonon wavenumber at rest is $\omega_G^0 = 1581 cm^{-1}$.[12,29,30] By manipulating equation (1) (see Supporting Information) we can easily obtain:

$$\varepsilon_{ll} = \frac{\Delta\omega_G^+ + \Delta\omega_G^-}{4\omega_G^0 \gamma_G} + \frac{\Delta\omega_G^+ - \Delta\omega_G^-}{2\omega_G^0 \beta_G} \;,\; \varepsilon_{tt} = \frac{\Delta\omega_G^+ + \Delta\omega_G^-}{4\omega_G^0 \gamma_G} - \frac{\Delta\omega_G^+ - \Delta\omega_G^-}{2\omega_G^0 \beta_G} \quad (2)$$

Since the values of both $\gamma_G$ and $\beta_G$ are now well established for graphene[18], it transpires that both the longitudinal and transverse strains can be calculated directly from the shift of the G sub-peaks at sub-micron steps across the specimen and thus avoiding any reliance on microscopic observations which are prone to errors due to the small values of strains present in the specimen. The Raman spectra (Figure 3) taken from both the suspended and the supported areas of the specimen show clearly that the graphene layer is 1LG with a well-defined 2D Lorentzian peak. Furthermore, the absence of the Raman PMMA peaks in spectra coming from graphene in air reveals the successful removal of the substrate. It is interesting to note that a weak D peak is observed in both the suspended as well as the embedded graphene. This may indicate the presence of defects that result from the e-beam irradiation in the suspended graphene, however, their presence also in the embedded graphene points as a likely source of defect generation the handling and exfoliation procedure and not the e-beam irradiation (although further investigation is necessary).

By measuring the wavenumber shifts Δω along the x-axis of the window of Figure 2 and for each incremental position along the y-axis we observe that the flake is subjected to a range of discrete axial strain values from ~0.4% to ~0.8%. The calculated values of the transverse strain $\varepsilon_{tt}$ showed that suspended graphene [18] is subjected to a range of lateral strains of 0.05% to 0.10%, which are far higher than the critical strain required for orthogonal buckling as discussed further below.

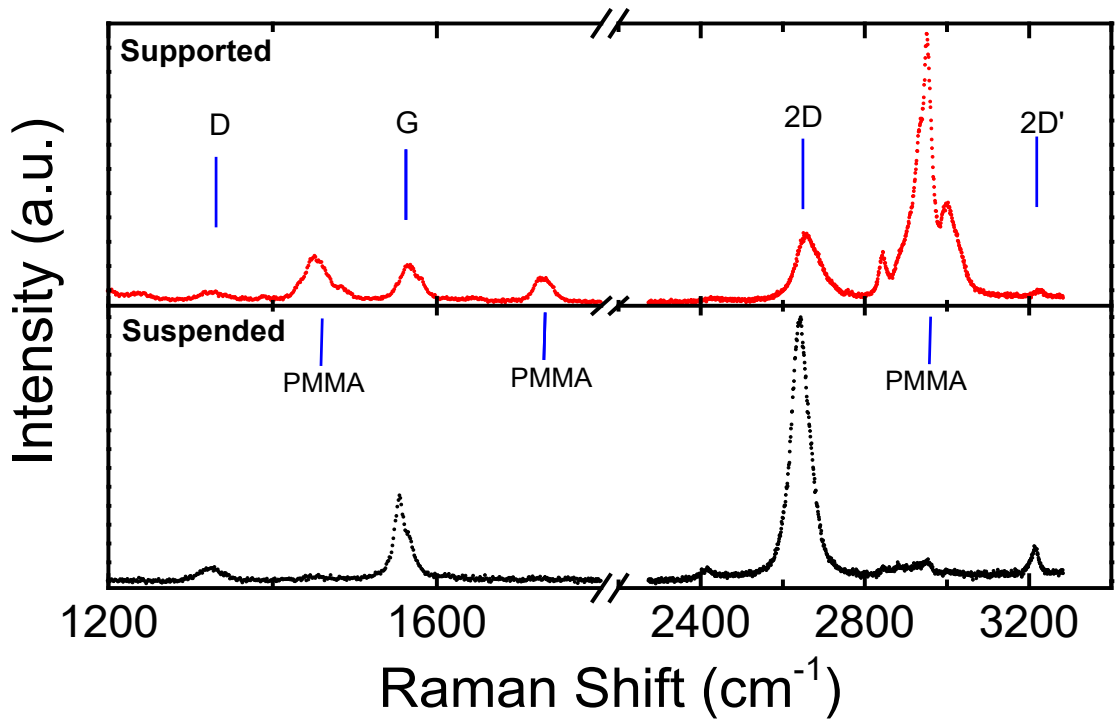

**Figure 3** Representative Raman Spectra of monolayer graphene taken from the supported and suspended regions. The G and 2D graphene peaks are clearly visible in both cases. Weak D and 2D' peaks are also shown. The PMMA peaks at ~1450 cm$^{-1}$, ~1730 cm$^{-1}$ and ~2952 cm$^{-1}$ are also visible in the supported region.

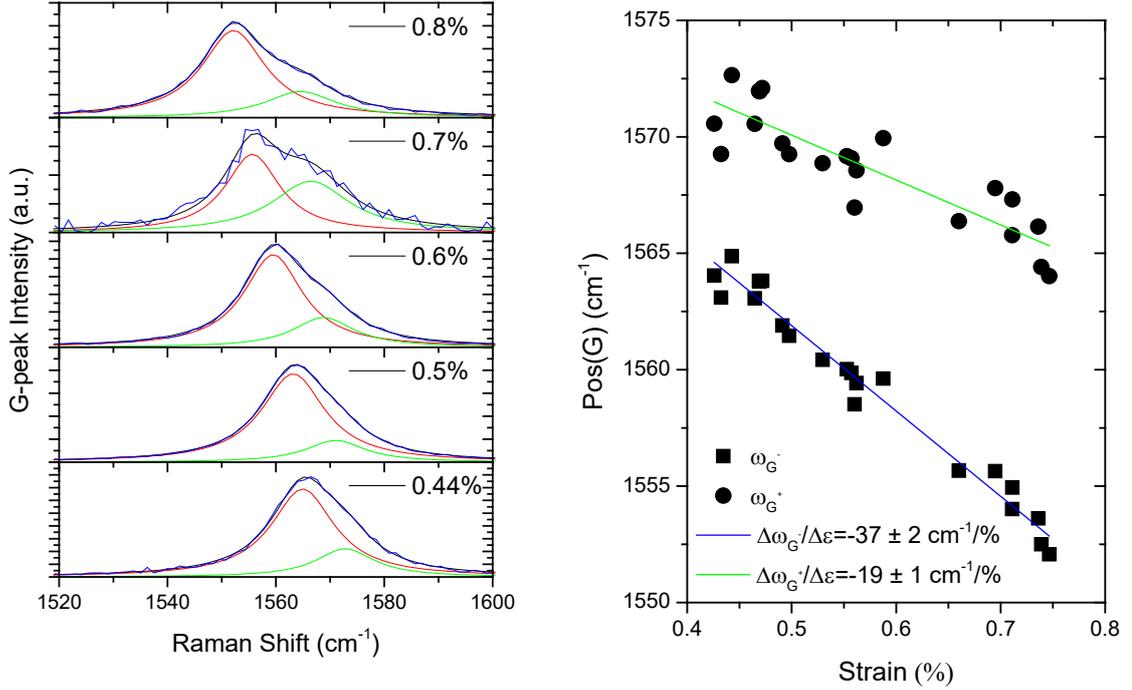

(a)  (b)

Figure 4 (a) Representative Raman spectra of the G-peak at various strain levels for the suspended 1LG; the splitting of the G$^-$ and G$^+$ components are clearly seen. Each strain level corresponds to a different lateral position across graphene. (b) G peak position (G$^-$ and G$^+$) as a function of strain. The straight lines are least-squares-fitted to the experimental data.

In Figures 4 and 5, we relate the strain derived from the phonon –secular- equations (1) and (2) with the values of G and 2D frequencies in each case. It is evident that for our analysis to be valid a linear relationship between frequency and strain should be obtained. As seen, the G-peak (Fig. 4) is clearly splitting into two components by the lifting of the $E_{2g}$ degeneracy at high strains since the two phonon eigenvectors are orthogonal (parallel and perpendicular to the strain direction). To the best of our knowledge, this is the first time that such a clear split of the G-peak in suspended graphene has been observed, which is in agreement with previously reported results for supported graphene. The least-squares-fitted straight lines to the experimental data exhibit slopes of -37±2 cm$^{-1}$/% and -19±1 cm$^{-1}$/% for the G$^-$ and G$^+$ components, which are very close to the predicted values of -36.4 cm$^{-1}$/% and -18.6 cm$^{-1}$/% obtained by Mohiudin et al [18] for graphene suspended in air. By comparing the relative intensities of the two components while keeping the strain axis parallel to the polarization of the excitation laser beam [18], it transpires that the graphene crystal is oriented at an angle $\varphi_s=17°\pm2°$ relative to the strain axis as shown on the upper left corner of Figure 2.

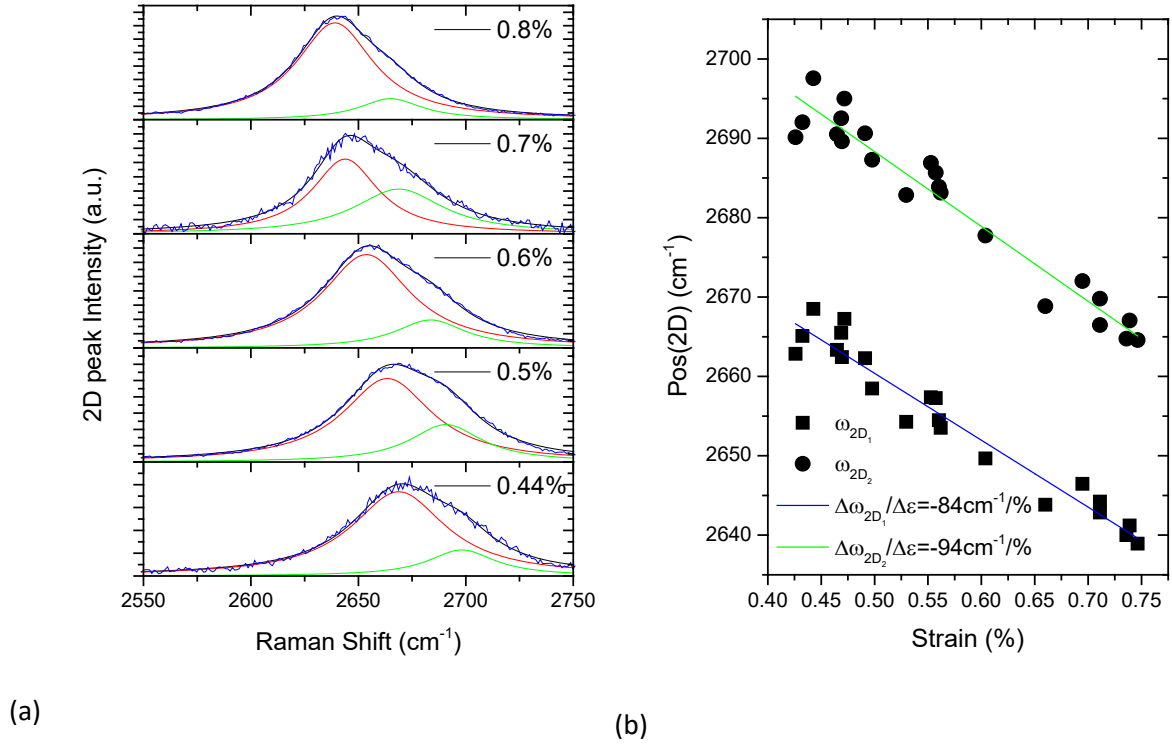

(a) (b)

**Figure 5 (a) Representative Raman spectra of the 2D-peak at various strain levels for the suspended 1LG; the splitting of 2D in air is clearly seen. Each strain level corresponds to a different lateral position across graphene. (b) 2D sub-peaks as a function of strain. The straight lines are least-squares-fitted to the experimental data.**

In Figure 5, representative spectra for the 2D peak are presented for the same range of axial strains. In this case a clear double peak [29,31,32] is observed which, as shown by Frank et al [16], is possibly due to the contribution of two distinct double resonance scattering processes (inner and outer) in the Raman signal. The splitting depends on the direction of the applied strain and the polarization of the incident light but also on the laser excitation line (strong effect at 785 nm). However, it is worth noting that for the suspended flake examined here at 514 nm excitation and for an orientation of $\varphi_s=17^o$ relative to the strain axis, a clear splitting of the 2D peak is also observed. The obtained strain sensitivities of -84 cm$^{-1}$/% and -94 cm$^{-1}$/% for the 2D$_1$ and 2D$_2$ peaks, respectively, are indeed extremely large and certainly correspond to the largest values of phonon shift ever recorded for uniaxial deformation. Again these values agree well with the predicted value of ~-83 cm$^{-1}$/% - [18] for free-hanging graphene based on results obtained from a simply-supported specimen.

In conclusion, the linear relationships within experimental error between G and 2D peak frequency values and axial strain derived from the secular equations (1) and (2), corroborates our initial premise of extracting the axial strain value at the sub-micron scale from the frequency shifts without the need of any other devices. However, as mentioned earlier, the calculation of axial strain for free lengths at the micrometre scale needs careful consideration. All attempts so far to stretch freely-suspended graphene [23,24] involved the gripping of flakes with polymer adhesives. However, due to the weak affinity of graphene to polymer matrices, such as PMMA, it is expected that the strain does not reduce to zero upon entry of graphene into the adhesive grips but,as expected from shear-lag principles. it diminishes to zero at some distance away from the edge of the suspended flake[33]. This introduces large errors into the strain calculations and may explain the confusing results obtained in the literature to date. Our strategy here is twofold; firstly, we calculate the strain through the shift of the Raman G peak along the strain direction (fig.2, x-axis) and for various lateral positions across the flake (fig.2, y-axis) at the centre of the suspended area. Secondly, we conduct Raman measurements through the PMMA layer in the grip region for both G and 2D peaks. The strain distribution along the line of maximum strain (~0.8%) in graphene, at the centre of the created arc in figure 2, is presented in Figures 6a, b for both G (a) and 2D (b) Raman peaks. The results are indeed quite revealing as they show that the portion of graphene which is sandwiched between the two PMMA layers is also strained. In fact, there is a parabolic decay of strain over a distance of 2.5 μm (left) and 1.5 μm (right) depending on the size of graphene that placed within the grips. The rather large decay length observed here is not surprising in view of the weak van der Waals bonding between graphene and PMMA. It is interesting to note here that in experiments on graphene simply supported on PMMA beams, transfer lengths of ~2 μm have also been measured at strains as low as 0.4% [33]. The balance of shear-to-normal forces[34] requires that the stress decay within the grips (PMMA) is given by:

$$\frac{d\sigma}{dx} = -\frac{2\tau_t}{nt_g} \quad \text{or} \quad \frac{d\varepsilon}{dx} = -\frac{2\tau_t}{nt_g E} \qquad (3)$$

where $\sigma$ is the axial stress acting on the flake, $\tau_t$ is the interfacial shear stress (ISS) between graphene (for both surfaces) and polymer, $n$ is the number of graphene layers (here $n=1$), $E=1$TPa is the modulus of graphene and $t_g =0.335$ nm is the thickness of the 1LG. Since the strain distribution is known we can obtain easily from eq. (3) the ISS distribution, $\tau_t$, within the grips. As seen, a maximum value of $\tau_t= 0.75$ MPa is obtained near the edges of the embedded portion of the flake (Figure 6 (c)). In fact experiments conducted by us [33] have shown that this value is very close to the upper ceiling of the ISS. One should expect that on further loading of the system, debonding (slipping) should occur. Based on these results we can now offer an explanation for the failure to accurately measure the stress and strain by conventional means at the nanoscale. In all reported cases[23,26], a normal force is applied to graphene by just pulling the polymer grips of a similar system to that studied here. The strain is measured through the displacement of the suspended length although the stress is applied to both embedded and suspended portions. However, as demonstrated here, the stress in the graphene is built within the polymer grips over distances equal or greater than the suspended length due to the weak van der Waals bonding between the two materials. In other words, if the transfer length at each end is small, as in the cases mentioned above, then effective stress transfer cannot be ensured. In such measurements the value of applied force (stress) is not necessarily the value transferred to the suspended length and, therefore, the stress-strain results are problematic. In contrast, the technique proposed here has the capability to measure strains in the specimen itself (suspended graphene) with sub-micron spatial resolution through the shift of the G phonons without resorting to optical observations of "grip" displacement. The only weakness of this technique is that it can only be applied to cases where stress is linearly related to strain, which- according to both numerical modelling[35] and experiment (AFM bending [36])- should be valid up to 10% strain.

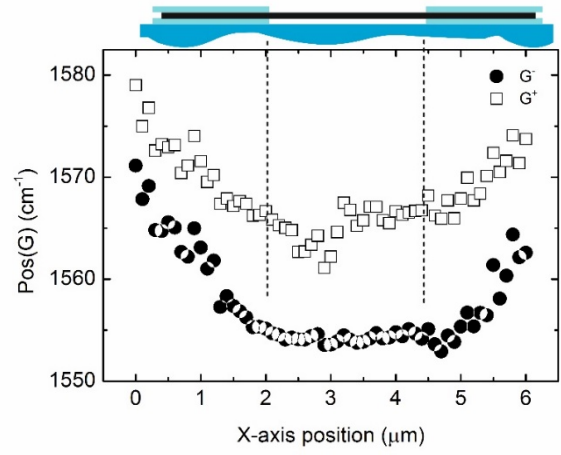

(a)

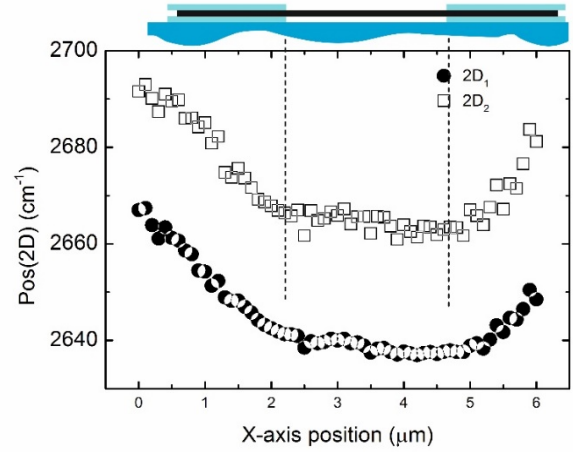

(b)

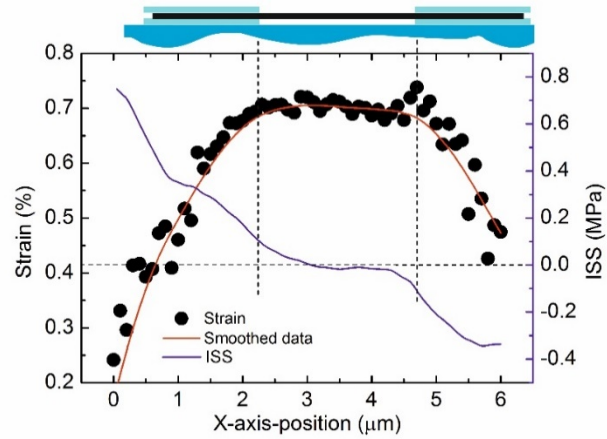

(c)

**Figure 6 (a) and (b) Raman shifts of G and 2D sub-peaks due to strain, along the graphene flake, both embedded in PMMA and in air. (c) Strain distribution (left axis) and ISS (right axis) along the graphene flake. Above each graph are depicted the embedded (sandwiched black line) and free-standing graphene (free black line) areas.**

We would like now to comment as to whether stress measurements are also possible by means of Raman[37] spectroscopy. In two recent publications, we have proposed the use of the stress sensitivity of either the G or 2D peaks[8,38] for independently converting the shift of the Raman wavenumber to values of stress. The idea here is based on the recorded relationship between the G or 2D strain sensitivity as a function of the Young's modulus of a number of graphitic materials such as carbon fibres and graphene. Since the relationship in both cases is linear, then the slope of the line that passes through the origin represents the stress sensitivity regardless of modulus. For the G peak the clear splitting into two peaks of quite different slopes make the calculations more cumbersome. However, for the 2D peak the slopes of the inner and outer components are not that different and the calculated average shift (see Supporting Information) of ~80 cm$^{-1}$/% can be easily employed for this purpose. For 514 nm excitation ($\omega_{2D}$ = 2680 cm$^{-1}$), the corresponding average value obtained from graphene loaded axially on flexible beams is -5.7 cm$^{-1}$ GPa$^{-1}$. For a maximum- axial- strain of 0.8% (Figure 6(c)) the corresponding maximum value of axial stress in the suspended part of the flake is ~9 GPa. [8,38]

We now turn our attention to possible out-of-plane deformations in graphene resulting from axial deformation along the suspended length of Figure 2. As mentioned earlier, due to the extremely small thickness of graphene (~0.335 nm) for an axial deformation of 0.8% the lateral compression strain is estimated to ~0.10% (for Poisson's ratio $\nu$=0.13 [18]), which is six orders of magnitude higher than the critical buckling strain of ~10$^{-9}$ (see Supporting Information). It is conceivable therefore that any suspended graphene membrane should be exhibiting orthogonal buckling under a tiny axial stress[39-43]. Here we provide strong indications that for 1LG loaded in air, small axial stress brings about orthogonal buckling and therefore out-of-plane effects are congruent to uniaxial deformation. In Figures 7 (a) and (b), we plot the intensity variation and the wavenumber values along the transverse direction (y-axis in fig. 2) to strain axis for all the G (a) and 2D (b) peaks of the suspended part of the flake. As seen, there is a periodic fluctuation of intensity of wavelength in the range of 0.6μm to 1μm. The pattern obtained is the same for all the components of both G and 2D peaks. These fluctuations cannot be attributed to interference effects with the glass substrate of the device and this is because of the simultaneous presence of similar fluctuations in the positions of the G and 2D peaks. Spectral shifts could not be originated from interference effects. The wavenumbers for all peaks seem to be decreasing as one moves from the left to the right-hand side of the graph (i.e. from low to high strain position). However, as mentioned earlier, this is expected (Figures 4 and 5) due to the decrease of the wavenumbers with tensile strain. However, careful examination of the data reveals that for each intensity trough there is a systematic local increase of G and 2D sub-peaks, as the applied axial strain increases. The opposite trend is observed for the positions of intensity crests for which a wavenumber decrease of equal magnitude is observed. In the case of G peak, wavenumber shifts of about 2 to 8 cm$^{-1}$ are observed, corresponding to a local strain variation of the order of 0.05 to 0.22%, which is not insignificant for a maximum applied strain of ~0.8%. The reason for this behaviour can be explained by the formation of a buckling wave in the transverse direction; the values near the crest and trough of the buckling wave are affected by the corresponding transverse tensile and compressive components and, thus, a systematic undulation of the wavenumber is clearly observed (Fig. 7). In other words the results depicted clearly in the Raman imaging of Fig. 8 point to the formation of an orthogonal buckling wave of a wavelength ranging from ~0.6 to ~1 μm.

It appears that at least for strains up to ~0.8%, the lateral collapse affects only marginally the imparted value of strain (and stress) that the material sustains. However, the formation of out-of-plane structures during axial deformation needs further examination possibly by SEM and AFM[43,44] as it may have important consequences at much higher strains for which it is conceivable that it affects the electronic properties of the material. It is interesting to note that due to the weak bonding between graphene and PMMA this buckling instability propagates into the portion of graphene that is embedded into the polymer grips (see Figure 8).

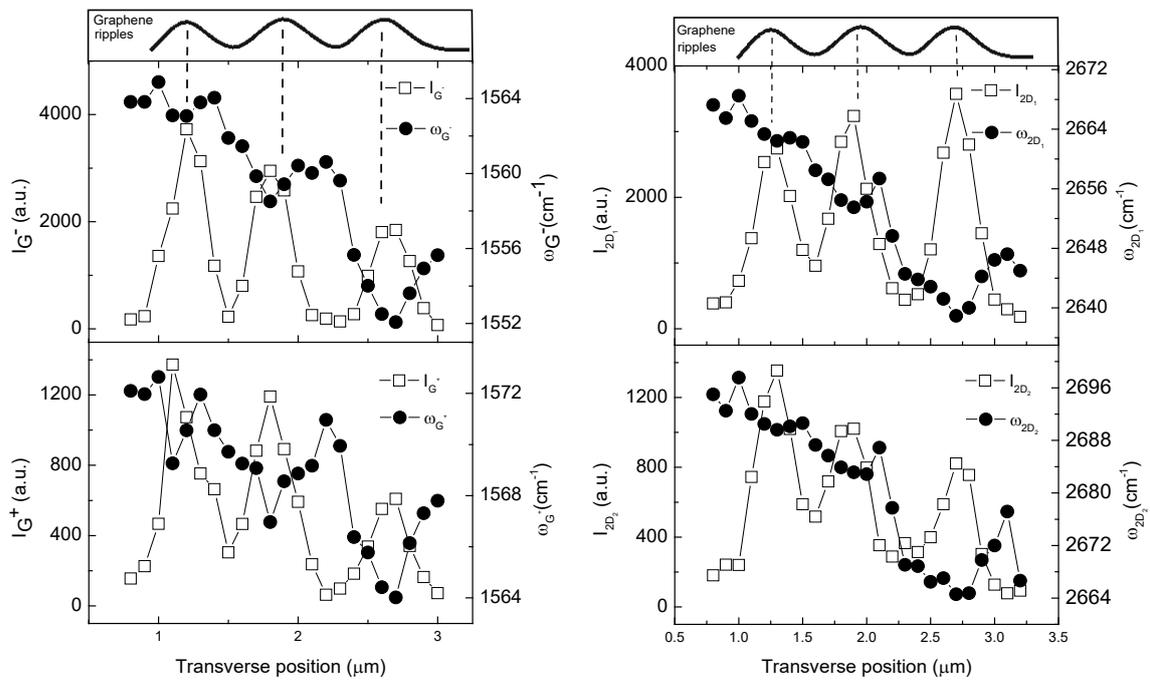

**Figure 7** Raman intensity and frequency evolution of the G and 2D sub-peaks in the transverse direction to strain axis (Y-axis in fig.1). Periodic fluctuations of both intensity and frequency are observed. This behaviour is attributed to the formation of an orthogonal buckling wave to graphene. The spatial resolution of the measurements was 100 nm.

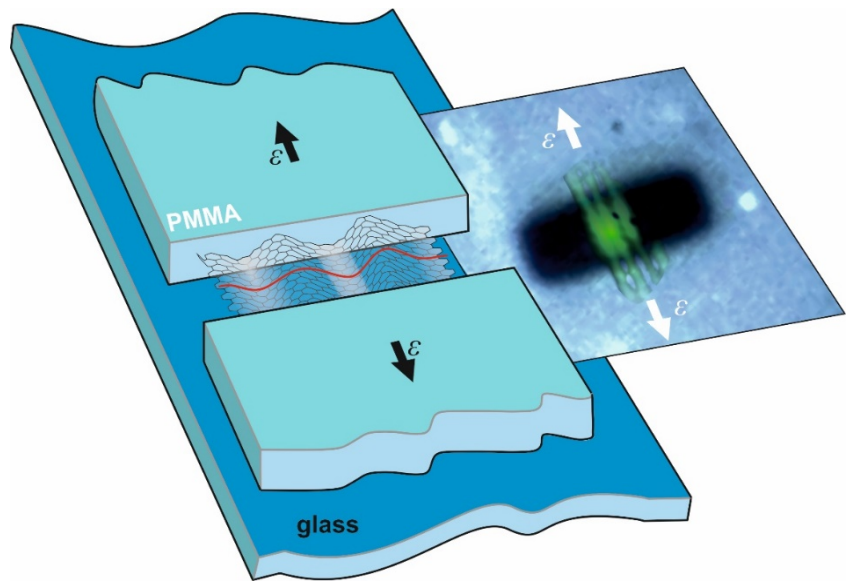

**Figure 8** Schematic of wrinkle (buckle) formation due to lateral compression and Raman map of the 2D peak intensity of graphene flake, I($\omega_{2D}$). A clear intensity undulation is shown in Raman map. The mapping step was 250 nm.

# Conclusions

We have presented true uniaxial measurements on one-atom thick 1LG freely suspended in air of gauge length of ~3 μm. This claim was supported by the huge phonon shifts observed in the strained graphene and the indication of lateral buckling as it is expected for any thin membrane stretched longitudinally in air. The specimen was prepared by sandwiching an exfoliated graphene flake into a PMMA polymer matrix and by removing a section of PMMA by e-beam lithography without damaging the graphene specimen (3-step process). Due to the fabrication procedure mechanical stress is accumulated on the sample and the graphene flake is subjected to a gradient of uniaxial strain. By employing a Raman microprobe and a nanomover translation stage we could map the increments of uniaxial strain across the flakes at a sub-micron resolution. The actual values of strain were measured in a non-destructive fashion through the large G-peak wavenumber shifts for each position following a procedure established earlier[12]. Clear splitting of both G and 2D peak was observed for the first time in suspended graphene. The Raman shifts of the split G and 2D peaks found to depend linearly on strain. The experimentally calculated slopes -for the first time for suspended graphene- are in accordance with predictions verifying the validity of our method. Measurements are also conducted on the portion of graphene embedded into the grips which revealed that the stress is transferred to the suspended part of the flake over large distances (>2 μm) due to the weak graphene/ PMMA interface; an interfacial shear stress of 0.75 MPa was measured at the edge of the flake which is close to the maximum value that can be sustained by such a system. This finding renders any conventional measurements attempted in the past uncertain, since the axial stress is in effect applied to a gauge length that extends into the polymer grips. Finally, we have provided evidence that axial loading of graphene, is always accompanied by the formation of orthogonal wrinkles similar to what is observed when a thin- macroscopic- membrane is stretched uniaxially[27,45]. The wrinkle formation causes a variation of local axial strain of the order of 0.05 to 0.22% which could induce premature failure at high strains. The work here exposes clearly all the problems encountered in the experimental mechanical measurements at the nanoscale and points to best practices when it comes to the application of uniaxial strain to 2D materials such as 1LG.

# *Methods*

## *Sample preparation*

Mechanically exfoliated graphene from Highly Oriented Pyrolytic. Graphite (HOPG) was deposited onto Si/300nm SiO$_2$ substrate. Graphene was identified initially by optical microscopy and afterwards by AFM. Then, a combination of e-beam lithography and oxygen reactive ion etching (RIE) was performed in order to isolate the single graphene flake (i.e. to remove adjacent multilayer graphite portions). Furthermore, several layers of PMMA (950k; 2.5%) were spin-coated on the samples in order to reach a thickness of ~1 μm. The graphene/PMMA sample, after being dipped in hot water at 90°C for 3 hours was manually transferred to another Si/300nm SiO$_2$ substrate with already deposited 1 μm thick PMMA layer. The entire sample was then baked for a few minutes at 160°C in order to remove the water trapped in between PMMA films. In this way, graphene was placed between two identical PMMA layers (forming a sandwich). Afterwards, a well-defined rectangular area (3μm x 10μm), across the graphene flake, was created by e-beam lithography (dose ~330 μC/cm$^2$). After this, the part of graphene flake inside the exposed area becomes suspended, (PMMA removed), while the parts at both sides of the opened window remain sandwiched between the PMMA layers. In this manner, graphene flake is clamped at both sides of the initially rectangular window (see Figure 1). Finally, the film was transferred on a release-agent covered glass substrate. The quality of suspended graphene, which depends on the successful removal of PMMA film above and below SLG, is verified accurately by RS.

Figure 3 shows representative spectra of the well-known G and 2D peaks of supported and suspended graphene taken from one of the investigated sample. In the case of supported graphene, two clear peaks, at the sides of G peak, at ~1450 cm$^{-1}$ and ~1730 cm$^{-1}$, are present and attributed to PMMA. Furthermore, next to the 2D peak of graphene a very intense PMMA peak is recorded at 2952 cm$^{-1}$. In the case of suspended graphene, these PMMA peaks vanish, revealing the absence of PMMA. It must be mentioned that at the centre of the opened window the laser spot covers only suspended areas of the graphene (the laser spot size is less than 1 μm for 514.5 nm excitation wavelength and the focusing geometry of out setup while the opened window is 3 μm wide) leading to a complete absence of PMMA peaks in Raman spectra. As the laser spot moves from the centre to the sides of the window the peaks that correspond to PMMA appear, although graphene is still suspended, and reach their maximum when the whole spot area is out of the window.

### *Raman Spectroscopy*

The experimental setup used for Raman characterization is the commercially available Renishaw, Invia Reflex 2000, MicroRaman system. A grating with 2400 lines / mm was used to resolve signal providing 2 cm$^{-1}$ spectral resolution. The samples were excited with the 514.5nm (2.41eV) line of an argon ion laser and the irradiation power was kept far below 1 mW to avoid local heating and sample destruction. This way, unwanted thermal spectral shifts and line-shape changes were avoided. A long working distance objective lens (100X, NA 0.85) from Leica was used to focus the laser light to a diffraction limited spot. Our measurements were performed under ambient conditions. In order to fully characterize the graphene sample with Raman spectroscopy extended mapping measurements were realized. The spatial resolution of our mapping stage was 100 nm.

## *Associated content*

### *Supporting Information.*

Microphotographs of sample preparation, Strain calculation from Raman Spectra, Converting spectroscopic data into stress- strain, Critical buckling strain

## *Author contributions*

The manuscript was written through contributions from all authors. All authors have given approval to the final version of the manuscript.

## *Acknowledgements*

Dr. Polyzos would like to acknowledge the support of the action "Supporting Postdoctoral Researchers" of the Operational Program "Education and Lifelong Learning", which is co-financed by the European Social Fund (ESF) and the Greek State through the General Secretariat for Research and Technology. The financial support of the European Research Council (ERC Advanced Grant 2013) no. 321124, "Tailor Graphene", is also gratefully acknowledged. Finally, FORTH/ ICEHT and Politecnico di Milano, acknowledge the financial support of the Graphene FET Flagship (''Graphene-Based Revolutions in ICT And Beyond''- Grant agreement no: 604391).

## *Figure captions*